# Diffraction limit of light in curved space

Jingxuan Zhang,[1] Chenni Xu,[1,2] Patrick Sebbah,[2] and Li-Gang Wang[1,*]

[1]School of Physics, Zhejiang University, Hangzhou 310058, China
[2]Department of Physics, The Jack and Pearl Resnick Institute for Advanced Technology, Bar-Ilan University, Ramat-Gan 5290002, Israel

*Corresponding author: lgwang@zju.edu.cn

Overcoming diffraction limit is crucial for obtaining high-resolution image and observing fine microstructure. With this conventional difficulty still puzzling us and the prosperous development of wave dynamics of light interacting with gravitational fields in recent years, how spatial curvature affect the diffraction limit is an attractive and important question. Here we investigate the issue of diffraction limit and optical resolution on two-dimensional curved spaces - surfaces of revolution (SORs) with constant or variable spatial curvature. We show that the diffraction limit decreases and resolution is improved on SORs with positive Gaussian curvature, opening a new avenue to super-resolution. The diffraction limit is also influenced by propagation direction, as well as the propagation distance in curved space with variable spatial curvature. These results provide a possible method to control optical resolution in curved space or equivalent waveguides with varying refractive index distribution and may allow one to detect the presence of non-uniform strong gravitational effect by probing locally the optical resolution.

## 1. Introduction

An inevitable problem in imaging process is that the resolution (the minimal distance between two points that can be discriminated) of conventional optical imaging systems is limited by diffraction, as a consequence of the loss in the far-field of evanescent waves carrying information on the high spatial-frequency components. The need to overcome the diffraction limit has led to the invention of the scanning near-field optical microscope (SNOM), which achieves subwavelength optical resolution[1] and inspired tremendous works in subwavelength imaging in recent years[2-8]. For instance, an intriguing idea to overcome diffraction limit is using the "superlens" made of silver, which was proposed by Pendry in 2000[2]. Later the idea was realized experimentally through planar left-handed lens[3] and material with negative permittivity or permeability[4]. Furthermore, the "hyperlens" was proposed and developed[5-7]. Besides, thanks to the prosperous development of laser technology for the decades, super-resolution fluorescence microscopy has obtained magnificent achievements[9-15], started with the stimulated emission depletion fluorescence microscopy proposed by Hell[9-10], followed by other proposals such as the reversible saturable optical linear fluorescence transitions11, the stochastic optical reconstruction microscopy[12], or the photoactivated localization microscopy[13]. However, all the works about overcoming diffraction limit have been considered in flat space without gravitational fields, so that the size of diffraction spot was only decided by the wavelength of light and the characteristics of the imaging system. It is well known from Einstein's general relativity (GR) that light follows curved-trajectories and unusual dynamics in the vicinity of black-hole. In this work, we investigate the effect of spatial curvature on diffraction limit, starting with Fraunhofer diffraction in two-dimensional (2D) curved space.

Wave dynamics of light in gravitational field has attracted much attention in recent years. Since measurable gravitational effect can be rarely obtained in laboratory, researchers have introduced several analog systems to simulate gravitational fields with laboratory devices[16-26], such as Bose-Einstein condensates[16-18], extremely low group velocity electromagnetic fluid[19], metamaterials with inhomogeneous refractive index distribution[20,21], or optical fibers[22]. Among these methods, one approach is to study the behavior of light in curved space by directly constructing spatial geometries that mimics distortion by gravitational field[27-42]. This field of research originates from Batz and Peschel's study of nonlinear Schrödinger equation and propagation of a Gaussian beam in 2D curved space[27], inspired itself by Costa's dynamics of particle constraint on surface[28]. Subsequently a series of theoretical[29-37] and experimental[38-42] investigations have been carried out, investigating Wolf effect in 2D curved space[29,30], spatial accelerating wave packets[31], Gouy phase shift[32], topological phase of photonic crystal[33], nonlinear dynamics[34-36], and, experimentally, the observation of self-focusing, diffraction-less propagation[38], accelerating wave packets[39], group and phase velocity[40], Hanbury Brown and Twiss effect[41] and surface plasmon polariton[42]. For the convenience and feasibility of calculation, the curved spatiotemporal geometries usually considered were often surfaces of revolution (SORs) embedded in the three-dimensional space, because of their rotational symmetry.

In this work, we take advantage of this paradigm to investigate the diffraction limit of light on SORs. Starting with Fraunhofer and wave equations, we analytically give the expression of intensity distribution of light diffracted by a slit on SORs with constant Gaussian curvature, and investigate the intensity distribution on the geodesic perpendicular to the propagation direction, as the mean to explore the diffraction limit, or equivalently the size of Fraunhofer diffraction spot. Then the effect of variable Gaussian curvature of curved space on diffraction limit is explored. When the propagation direction is not along the longitude of SORs, we calculate the intensity distribution of diffraction field through Huygens principle in curved space. This method allows us to study propagation with arbitrary initial position and direction, instead of simply taking the equator as input plane and longitude as propagation direction. We also apply this method to other surfaces with variable Gaussian curvature. Here we consider different propagation distances and directions to calculate the diffraction intensity distribution, and compare them with the case of flat space. We demonstrate that the positive spatial curvature decreases the diffraction effects to get a better optical resolution, which gives a new method to super-resolution. Moreover, the non-uniform distribution of spatial curvature varies the optical resolution in different directions, indicating that the anisotropy of space can be investigated by observing the variation of optical resolution in different directions. The results presented here may also stimulate imaging technologies in planar waveguides with transformation optics and stir up electron optics in 2D curved material.

## 2. Diffraction on SORs with constant Gaussian curvature

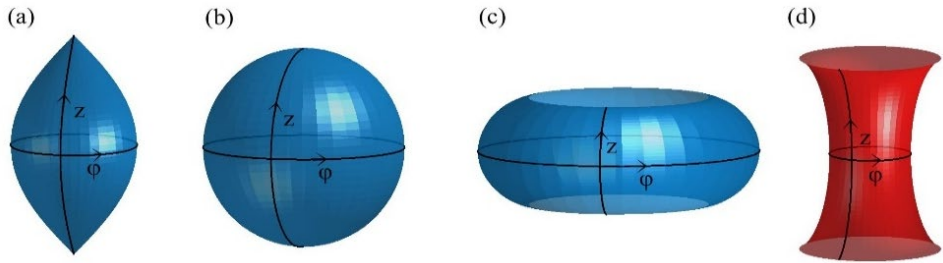

**Fig. 1.** SORs with constant positive (a)-(c) and negative (d) Gaussian curvature. Here (a) spindle with $r_0<R$, (b) sphere with $r_0=R$, and (c) bulge with $r_0>R$.

Firstly, we consider a family of SORs with constant Gaussian curvature (CGC) (see Fig. 1), which are described by the metric:

$$ds^2 = dz^2 r_0^2 \cos_q^2(z/R)d\varphi^2, \tag{1}$$

where $z$ is the proper length along the longitudinal direction, $\varphi$ is the rotational angle, $r_0$ is an initial rotational radius at $z=0$, $R = |K|^{-\frac{1}{2}}$ is the radius of a SOR with Gaussian curvature $K$, $q=\text{sgn}(K)$, and $\cos_q(x)$ is defined as $\cos(x)$ for $q=1$ (i.e. $K>0$), and $\cosh(x)$ for $q=-1$ ($K<0$). When $K>0$, the range of $z$ is $(-\pi R/2, \pi R/2)$ for $R \geq r_0$, and it becomes $(-R\arcsin(R/r_0), R\arcsin(R/r_0))$ for $R<r_0$. When $K<0$, $z \in (-R\text{arcsinh}(R/r_0), R\text{arcsinh}(R/r_0))$. The propagation of a light field in curved space satisfies the scalar wave equation[38]:

$$\Delta_g U + (k^2 + H^2 - K)U = 0, \tag{2}$$

where $\Delta_g = \partial_i(\sqrt{g}g^{ij}\partial_j)/\sqrt{g}$ is the covariant Laplace operator, $g$ is the determinant of metric tensor matrix $\mathbf{g}$, $g^{ij}$ is the inverse matrix element of $\mathbf{g}$, $H$ is the extrinsic (or mean) curvature which is decided by the topology of curved space, and $k$ is the wavenumber of light. In this work, we consider a series of SORs which possess significantly small extrinsic curvature $H$ and Gaussian curvature $K$ compared with wavenumber of light $k$, thus the term $H^2 - K$ can be neglected. By using the metric Eq. (1), the wave equation can be written as:

$$\frac{\partial^2 U}{\partial z^2} - \frac{q\tan_q(z/R)}{R}\frac{\partial U}{\partial z} + [r_0\cos_q(z/R)]^{-2}\frac{\partial^2 U}{\partial \varphi^2} + k^2 U = 0. \tag{3}$$

Now, one can obtain the point spread function (PSF) of 2D curved space by solving Eq. (3) under paraxial approximation[41],

$$h_q(\varphi, \theta, z) = \sqrt{\frac{1}{i\lambda R \sin_q(z/R)}} \exp\left[ikz + \frac{i}{2k}\int_0^z V_{\text{eff},q}(z')dz' + \frac{ikr_0^2(\varphi-\theta)^2}{2R\tan_q(z/R)}\right], \tag{4}$$

where $\lambda$ is the wavelength of light, and $\varphi, \theta \in [-\pi, \pi]$ are the rotational angles at $z=0$ and $z>0$, respectively. $V_{\text{eff},q} = \frac{q[1+\cos_q^{-2}(\frac{z}{R})]}{4R^2}$ is the effective potential caused by the spatial curvature. To investigate the optical spatial resolution in curved space, we start from Fraunhofer far-field diffraction. At the equator ($z=0$), we place a slit with width $2a$, whose center is the origin of coordinate, and consider a plane wave with amplitude $U(z=0)=1$ propagating along the longitude towards the slit. The intensity of the diffraction field can be obtained by

$$I_q^s(\theta, z) = \left|\int_{-\frac{a}{r_0}}^{\frac{a}{r_0}} U(z=0)h_q(\varphi, \theta, z)r_0 d\varphi\right|^2$$
$$= \frac{1}{R\sin_q(z/R)}\text{sinc}^2\left[\frac{kar_0\theta}{R\tan_q(z/R)}\right]. \tag{5}$$

Here, we neglect the quadratic term of $\varphi$ under the Fraunhofer far-field approximation. Similarly, in the case of double-slit with width $2a$ and distance $d$, the intensity distribution is given by

$$I_q^d(\theta, z) = \frac{1}{R\sin_q(z/R)} \text{sinc}^2\left[\frac{kar_0\theta}{R\tan_q(z/R)}\right] \cos^2\left[\frac{kr_0\theta(a+d/2)}{R\tan_q(z/R)}\right]. \tag{6}$$

Clearly, from Eq. (5), the width of the center fringe of single-slit diffraction is fully determined by the zeros of the sinc function, while double-slit interference pattern from Eq. (6) is modulated by the sinc function and its fine structure is further modulated by the cosine function. However, when the propagation of light fields is considered in curved space, its intensity distribution is must be considered along a geodesic (the equivalent of a straight line in flat space) perpendicular to the direction of propagation[37]. Thus, in order to obtain the diffraction pattern along a geodesic in curved space, we need to know the geodesics perpendicular to the propagation direction and then calculate the intensity of each point on the corresponding geodesics to observe.

It is known that the geodesic in curved space satisfies the equation[37]

$$\frac{d^2 x^\sigma}{ds^2} + \Gamma^\sigma_{\mu\nu} \frac{dx^\mu}{ds} \frac{dx^\nu}{ds} = 0, \tag{7}$$

where $\Gamma^\sigma_{\mu\nu} = \frac{1}{2} g^{\sigma\rho} \left( \frac{\partial g_{\rho\mu}}{\partial x^\nu} + \frac{\partial g_{\rho\nu}}{\partial x^\mu} - \frac{\partial g_{\mu\nu}}{\partial x^\rho} \right)$ is Christoffel connection. Substituting Eq. (1) into Eq. (7), we obtain the geodesic[37]

$$d\varphi = \pm \frac{\kappa_q}{r_0 \cos_q(z/R) \sqrt{r_0^2 \cos_q^2(z/R) - \kappa_q^2}} dz, \tag{8}$$

where $\kappa_q = \left[ r_0^2 \cos_q^2(z/R) \frac{d\varphi}{ds} \right]_{\text{initial}}$ is a constant determined by the initial condition. For any on-axis propagation position ($z_0$, $\varphi_0$), the output plane composed of the geodesics that is perpendicular to the optical axis can be described by the points of ($z$, $\varphi$). In practice, one should first obtain the proper length $x$ for a given $z$ by substituting Eq. (8) into Eq. (1) and integrating that equation from $z_0$ to $z$. The proper length $x$ for the SORs described by Eq. (1) can be expressed as

$$x(z) = R\arcsin\left[\frac{r_0 \sin(z/R)}{\sqrt{r_0^2 - \kappa_1^2}}\right] - R\arcsin\left[\frac{r_0 \sin(z_0/R)}{\sqrt{r_0^2 - \kappa_1^2}}\right], \tag{9}$$

for $K>0$, while for $K<0$, it becomes

$$x(z) = \ln\left|\frac{\sinh(z/R) + \sqrt{\sinh^2(z/R) + 1 - \kappa_{-1}^2/r_0^2}}{\sinh(z_0/R) + \sqrt{\sinh^2(z_0/R) + 1 - \kappa_{-1}^2/r_0^2}}\right|. \tag{10}$$

Using the above equations (9) and (10), for a given proper length $x$ distanced from the on-axis point ($z_0$, $\varphi_0$), in turn we can obtain the value of the coordinate $z$. Then using Eq. (8), we finally obtain the coordinate information of every point ($z$, $\varphi$) on the output plane along the geodesics (perpendicular to the optical axis). For the SORs with $K>0$, we have

$$\pm(\varphi - \varphi_0) = \frac{R}{r_0}\arcsin\left[\frac{\kappa_1 \tan(z/R)}{\sqrt{r_0^2 - \kappa_1^2}}\right] - \frac{R}{r_0}\arcsin\left[\frac{\kappa_1 \tan(z_0/R)}{\sqrt{r_0^2 - \kappa_1^2}}\right], \tag{11}$$

while for $K<0$, we have

$$\pm(\varphi - \varphi_0) = \frac{R}{r_0} \ln \left[ \frac{\tanh(z/R) + \sqrt{\tanh^2(z/R) + r_0^2/\kappa_{-1}^2 - 1}}{\tanh(z_0/R) + \sqrt{\tanh^2(z_0/R) + r_0^2/\kappa_{-1}^2 - 1}} \right]. \tag{12}$$

Since the other three coordinates $z_0$, $\varphi_0$ and $z$ are known, $\varphi$ can be obtained from Eq. (11) or (12) directly. Substituting the coordinates $z$ and $\varphi$ into Eq. (5) and (6), we obtain the intensity of a point with proper length x on the output plane. Note that the sign of every proper length is defined by the sign of $\pm(\varphi - \varphi_0)$. In this way, we can calculate the intensities of all the points with different $x$, finally the intensity distribution along the different output planes can be obtained.

Figures 2(a1) and 2(a2), respectively, show the intensity distributions of diffraction through a single-slit and of interference through a double-silt at distance $z=300$ mm on SORs with positive, negative and zero curvature (flat space). Compared with the cases of flat space, the widths of the main-spot peaks (i.e., the main-spot size of the central fringes for both single-slit and double-slit cases) in the cases of $K>0$ are smaller, which means optical resolution has been enhanced compare to flat space. In contrast, on the SORs with $K<0$, the width of the main-spot becomes larger, indicating a degraded resolution. To investigate the evolution of the width of the central fringe as light propagates along the longitude, we consider the interval $\Delta x^{s,d}$ between the first two zeros of the sinc function around the central fringe. Here, the superscripts "$s$" and "$d$" denote single-slit and double-slit cases, respectively. Figures 2(b1) and (b2) demonstrate the impact of the spatial curvature on the evolution of $\Delta x^{s,d}$ with increasing distance. We can see that the width of the central fringe for both single-slit and double-slit cases in the curved space with positive $K$ is always smaller than that in flat space. This effect is enhanced as the propagation distance increases.

Figures 2(c) and 2(d) display the evolutions of diffraction and interference light fields in different spaces. It is seen that in flat space, the diffraction and interference fields propagate and expand along straight lines, however the diffraction process of light fields is suppressed/enhanced in curved spaces with constant positive/negative Gaussian curvature. These evolution features correspond to the narrowing or expanding effect of the central fringe width in the curved spaces with positive or negative values of $K$, compared with the case in flat space.

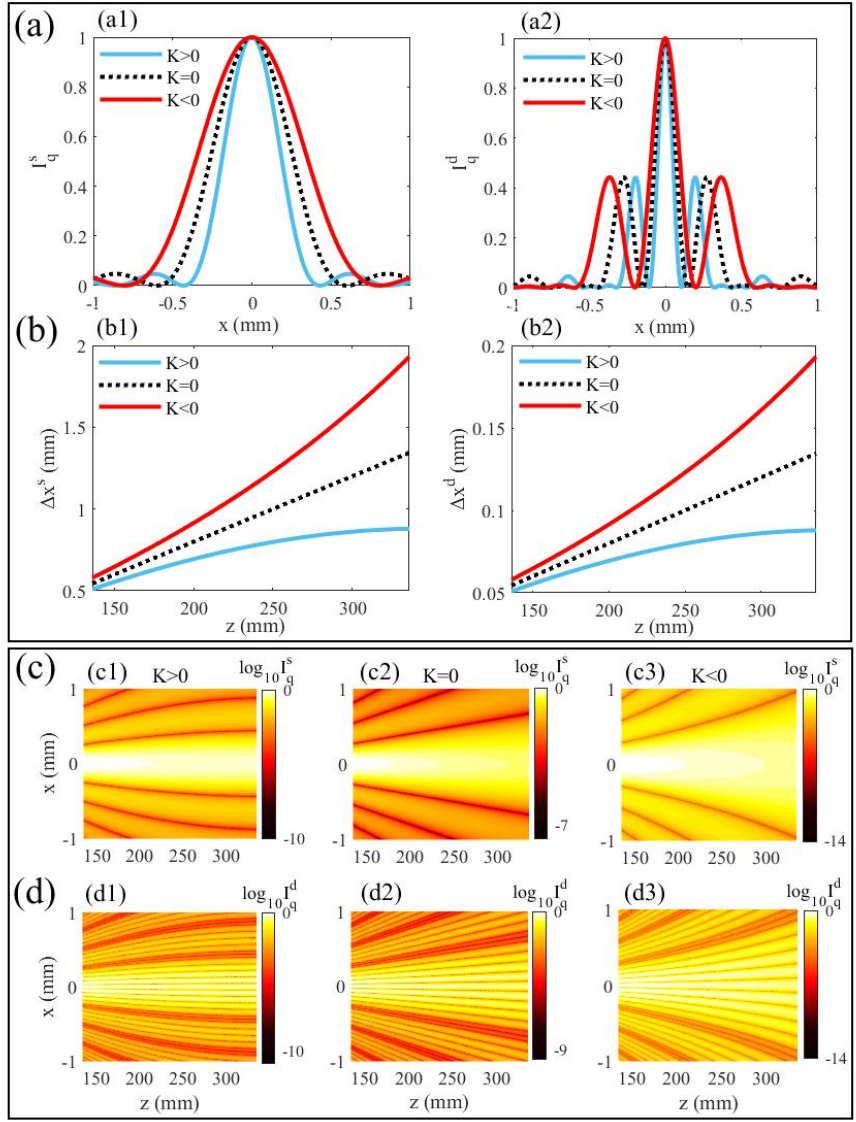

**Fig. 2.** Diffraction and interference on curved space. (a) Intensity distributions of Fraunhofer single-slit diffraction (a1) and double-slit interference (a2) of light at $z$=300 mm on SORs with different Gaussian curvature. (b) Variations of the central fringe widths $\Delta x^s$ and $\Delta x^d$ for single-slit diffraction (b1) and double-slit interference (b2) with the propagation distance in different SORs. (c-d) Evolutions of light fields of Fraunhofer single-slit diffraction (c) and double-slit interference (d) in different spaces: (c1, d1) $K$>0, (c2, d2) $K$=0, and (c3, d3) $K$<0. Here the parameters for SORs with $R$=220 mm are taken as $K$=20.66 m$^{-2}$ for $K$>0, and $K$=−20.66 m$^{-2}$ for $K$<0. Other parameters are $\lambda$=400 nm, $r_0$=100 mm, $a$=0.1 mm, $d$=0.2 mm in (a2) and $d$=0.8 mm in (d1)-(d3) for better visualization.

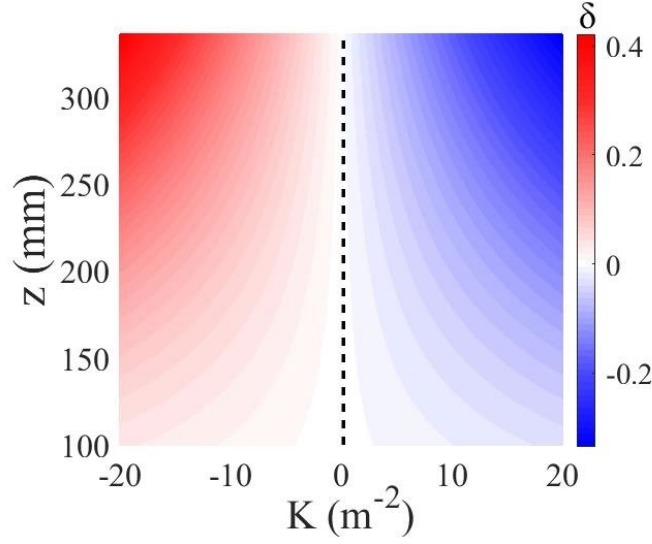

**Fig. 3.** The effect of Gaussian curvature $K$ of SORs on the change of the diffraction limit at different propagation distance $z$. The black dashed line denotes the case for the diffraction limit in flat space. The zero value of $\delta$ means no effect on diffraction limit, comparing with the case in flat space.

For conveniently discussing the influence of spatial curvature on the central fringe widths of diffraction pattern in far-field regions, we define a relative quantity, $\delta = \frac{\Delta x - \Delta x_f}{\Delta x_f}$, to quantify the narrowing or expanding effect of the central fringe width (i.e., the main-spot width) of light fields as the change of the diffraction limit in curved space. Here $\Delta x$ and $\Delta x_f$ are the widths for the central peak in curved and flat space, respectively. Clearly, $\delta < 0$ indicates that the optical resolution has improved, whereas for $\delta > 0$ it has degraded. Figure 3 shows the non-uniform dependence of $\delta$ with propagation distance and spatial curvature. The large positive/negative Gaussian curvature and long propagation distance will decrease/increase the diffraction limit apparently. Thus, one may expect a method to control optical resolution and diffraction limit by varying spatial curvature of non-Euclidian spaces.

However, the analytical solution we obtained above is not applicable when light propagation is not along the longitude or the input plane is not along the equator anymore. There is another theory based on the Huygens principle to get the solution for the general situation of light propagation on arbitrary curved surfaces along arbitrary propagation direction[37]. According to Ref. [37], the complex amplitude of light field at the output plane on 2D curved surface can be expressed by

$$U_{\text{output}}(P) = \sqrt{\frac{1}{i\lambda}} \int_l U_{\text{input}}(P_0) \frac{e^{ikL(P_0,P)}}{L(P_0,P)} A(P_0,P) dl, \qquad (13)$$

where $U_{\text{output}}(P)$ is the amplitude on the output plane, $U_{\text{input}}(P_0)$ is the amplitude on the input plane which can be considered as constant under far field approximation, $A(P_0, P)$ is the obliquity factor, which can be taken as unity when the propagation distance is long enough (i.e., one order of magnitude larger than the transverse dimension on the input plane[37]), and $L(P_0, P)$ is the eikonal function of light and is the geodesic length between two points on the input and output plane, which can be calculated by Eq. (8) for SORs described by Eq. (1). In practice, we set the input plane located at the object plane (in our case, we use single slit as the objects), meanwhile the output plane is always chosen the plane along a geodesic perpendicular to the propagation direction. The intensity distribution of light fields after a slit on curved space can be obtained by numerically calculating thousands of geodesics connecting the input and output planes according to the procedure described in Ref. [37]. Apparently, although this method does not allow to obtain the analytical solution, it has no limit for the choice of incident plane, propagation direction and distance. It is therefore an alternative to investigate light propagation in curved spaces.

Figure 4 further demonstrates the intensity evolution of light via a single-slit diffraction for three kinds of Constant-Gaussian-curvature SORs, when the incident plane is at a different angle with the longitude direction (i.e., the propagation direction is at an angle of $\Theta$ to the longitude direction). Clearly, although the propagation direction is not along the longitude direction anymore, the diffracting fields will focus in curved space with positive Gaussian curvature. Propagation direction does not affect at all the diffraction intensity distribution of light beam when Gaussian curvature is constant [Note that the slight difference in intensity patterns is presented due to the calculation errors of the integral in Eq. (13)]. The effects induced by spatial curvature are more apparent at longer propagation distance.

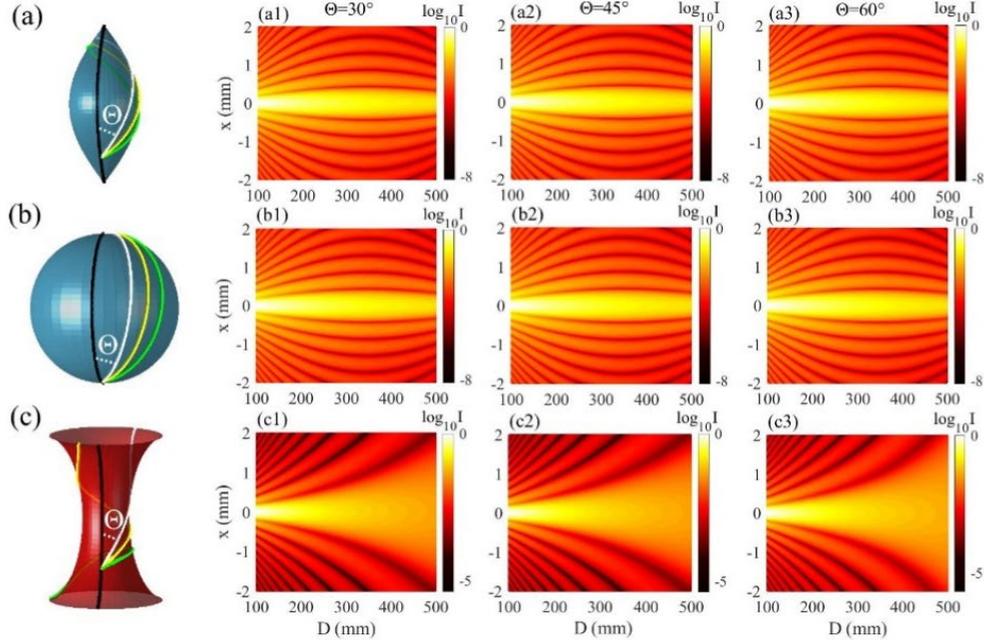

**Fig. 4.** Diffraction of light fields along different propagation directions on (a) spindle, (b) sphere and (c) hyperboloid. The propagation direction is described by the angle $\Theta$ between the longitude (black solid lines) and the incident direction. Here three typical propagation directions with $\Theta$=30°, 45° and 60° are considered, and the corresponding geodesics are indicated by the white, yellow and green curves. $D$ is the propagation distance along these geodesics. The initial central position of the input plane is $z_i$=−230.384 mm, −295.134 mm and −166.415 mm in (a), (b) and (c), respectively; and the parameters of these SORs are $r_0$=100 mm, 220 mm, 100 mm in (a), (b) and (c), respectively; and other parameters are same as in Fig. 2.

## 3. Diffraction on other typical SORs with varying spatial curvature

### 3.1 Two-dimensional Schwarzschild metric

Now let us consider the diffraction effect of light in a typical model, such as a Schwarzschild black hole, which is a typical solution of Einstein field equation, used for mimicking a light bending near a black hole. In such space, the spatial curvature changes as spatial position varies.

The spacetime metric of a Schwarzschild black hole is characterized by the line element as follows[43]

$$ds^2 = -\left(1-\frac{r_s}{r}\right)c^2 dt^2 + \left(1-\frac{r_s}{r}\right)^{-1} dr^2 + r^2\sin^2\Psi d\varphi^2 + r^2 d\Psi^2, \quad (14)$$

where $c$ is the speed of light, $r_s$ is the Schwarzschild radius, $\Psi$ and $\varphi$ are the polar and azimuthal angles, respectively. Due to the spherical symmetry of Eq. (14), we here only consider the propagation of light diffraction along its Equatorial plane. Thus, we take $\Psi=\pi/2$ for convenience without loss of generality. Then the spatial part of Eq. (14) becomes a Flamm's paraboloid (FP)[40] when omitting the temporal term, and the metric becomes

$$ds^2 = \left(1-\frac{r_s}{r}\right)^{-1} dr^2 + r^2 d\varphi^2, \quad (15)$$

with negative and variable Gaussian curvature $K(r) = -\frac{r_s}{2r^3}$. For an arbitrary SOR, its metric can be rewritten as $ds^2 = \left[1+\left(\frac{dH}{dr}\right)^2\right]dr^2 + r^2 d\varphi^2,$ where $H$ can be seen as the height of a point on the SOR, the geodesic equation can be also analytically obtained as[37] $d\varphi = \pm\frac{\kappa}{r\sqrt{r^2-\kappa^2}}\sqrt{1+\left(\frac{dH}{dr}\right)^2}\,dr$ with $\kappa = \left[r^2\left(\frac{d\varphi}{ds}\right)\right]_{initial}$ is a constant determined by the initial condition. In the current case, here $H(r) = \pm 2\sqrt{r_s(r-r_s)}$ for $r > r_s$. When we consider the diffracting fields propagating from far to near the event horizon of Schwarzschild black hole, as seen in Fig. 5(a), the evolution of the diffraction patterns depends on both the propagation direction and the propagation distance. From Fig. 5(a), we can see that when the propagation distance is far away from the event horizon, where the Gaussian curvature $K(r)$ is very small so that the space is close to flat space, thus the evolution of the diffraction field is initially like the case in flat space, while it diverges as the field approaches the black hole, since the spatial curvature becomes negatively large. Thus, we see that the width of the diffraction central fringe is much wider than the case in flat space due to stronger negative spatial curvature near the black hole. As the curvature is negative everywhere on FP, in principle the diffraction limit near the space of a black hole is always larger than that in flat space.

### 3.2 Two dimensional Schwarzschild-de Sitter (SdS$_2$) metric
It is well known that the spacetime near a black hole may be modified due to the presence of some matter-energy distributions[43]. One possible situation is that a black hole may be immersed in the background of dark energy which is characterized by a cosmological constant ($\Lambda$). Then Eqs. (14) and (15) are modified with the existence of positive cosmological constant $\Lambda$ in the above subsection (3.1), the metric becomes the Schwarzschild-de Sitter metric. By similarly taking a 2D slice, the line element is given by:

$$ds^2 = \left(1-\frac{r_s}{r}-\frac{\Lambda r^2}{3}\right)^{-1} dr^2 + r^2 d\varphi^2, \quad (16)$$

which has two horizons. Now the Gaussian curvature of such SOR becomes $K(r) = \frac{\Lambda}{3} - \frac{r_s}{2r^3}$, which is negative near black-hole horizon and become positive near cosmological horizon. Similar to the above considerations, if the input plane we set is near the cosmological horizon and the light propagates toward the black-hole horizon, the diffraction limit is initially smaller than that in flat space when the propagation distance is short at first, where the space is de-Sitter-domain and the spatial curvature is almost positive. As shown in Fig. 5(b), the diffraction patterns are bending towards the optical axis in the short regions. When light propagates into the space of Schwarzschild-domain, curvature goes negative and the width of main-spot peak increases rapidly, which increases the diffraction limit, also see Fig. 5(b1). Thus, we can find that in the de-Sitter-domain regions optical diffraction limit becomes smaller, indicating a better resolution of optical imaging, while optical diffraction limit increases quickly (i.e., a worse imaging resolution) in the Schwarzschild-domain regions.

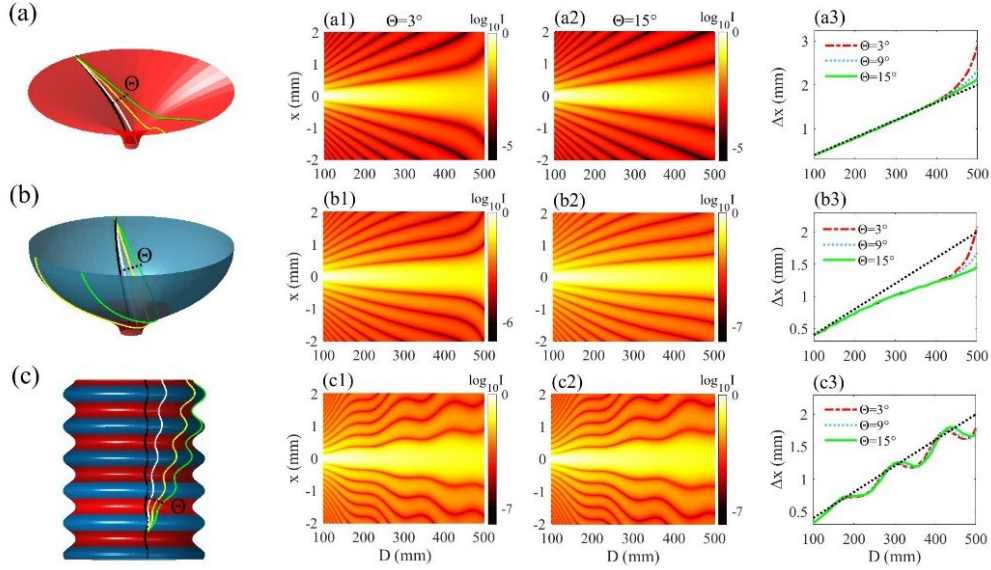

**Fig. 5.** Diffraction of light fields along different propagation directions on (a) a FP surface, (b) a SdS$_2$ surface and (c) a PPSS surface. The propagation direction $\Theta$ is defined as the same as in Fig. 4, and three typical directions $\Theta=3°$, $9°$ and $15°$, corresponding to the white, yellow and green geodesics, respectively, are plotted. The surfaces are colored by blue/red to indicate the regions with positive/negative Gaussian curvature. The black dotted lines in (a3), (b3) and (c3) show the variation of diffraction limit in flat space. Here the initial central positions of the input plane are $r_i$=460 mm in (a), $r_i$=283.66 mm in (b) and $z_i$=100 mm in (c), the Schwarzschild radius $r_s$=30mm in (a) and (b), the cosmological constant is $\Lambda$=33.33 m$^{-2}$ in (b). In (c), we take $A$=100 mm, $B$=10 mm, $\beta$=20 mm, $\varphi_0$=−1.25$\pi$. Other parameters are the same as in Fig. 2.

### 3.3 Diffraction on the surface with periodic peanut-shell shape (PPSS)

In this subsection, a special SOR with periodic structures (see Fig. 5(c)), called periodic peanut-shell shape (PPSS)[30,44], is introduced out of our interest. For convenience, we apply longitudinal proper length $z$ and rotational angle $\varphi$ coordinates system $(z, \varphi)$ to describe this surface. The metric of PPSS is given by[30]

$$ds^2 = dz^2 + \left[A - B\cos\left(\frac{z}{\beta} + \varphi_0\right)\right]^2 d\varphi^2, \tag{17}$$

where $A$ is the average rotational radius, $B$ is the amplitude of the periodically variable radius, $\beta$ is the period and $\varphi_0$ is the initial phase. On this SOR, the Gaussian curvature is given by $K(z) = \left[1 - \frac{A}{A - B\cos\left(\frac{z}{\beta} + \varphi_0\right)}\right]/\beta^2$, which has the oscillating property depending on the longitudinal proper length $z$ and may lead to positive and negative spatial curvatures. Thus, according to the above discussion, it is expected that the diffraction limit on such PPSS also reflects the oscillation behavior. Figure 5(c) shows the wavy feature of diffracting light fields with the increasing propagation distance. We observe that the diffraction field may rapidly expand in certain regions while present converging feature in other regions, respectively, indicating increased or decreased optical resolution, because the distribution of Gaussian curvature on PPSS is alternatively changing between positive and negative. Meanwhile, the spatial width of the diffraction central fringe is also found to depend on both the propagation direction and distance as shown in Fig. 5(c3).

Finally, in Fig. 6, we plot how the propagation direction on different SORs affects the diffraction limit at different propagation distances. It is found that on the SORs with constant Gaussian curvature, the diffraction limit does not vary with propagation direction, at fixed propagation distance. Clearly, as discussed above, the diffraction limit is smaller for positive $K$ and larger for negative $K$, compared with the flat-space case. However, the relative quantity, $\delta$, changes as with propagation direction for other families of SORs with non-constant spatial curvature. On the surface of FP, it is seen that as the angle $\Theta$ increases, the diffraction limit tends to decrease and approach its value in flat space since the propagation of the light beam gradually deviates from the Schwarzschild black hole. Meanwhile, the propagation distance also impacts the diffraction limit. On the SORs of $SdS_2$ metric, the change of the diffraction limit as a function of the propagation direction is similar to that on FP, but the value of the related quantity $\delta$ can change from positive to negative and the sign of $\delta$ is dependent on the local spatial curvature. Similarly, on PPSS, the relative quantity $\delta$ oscillates as $\Theta$ grows, which indicates how optical resolution deteriorates or improves compared with that of flat space. From our result, since the non-uniform distribution of spatial curvature varies the optical resolution, the isotropy and anisotropy of space may be reflected by the observation of the variation of optical resolution. Thus, we may detect the presence of anisotropic universe and spacetime by observing the variation of optical resolution in different propagation directions and distances.

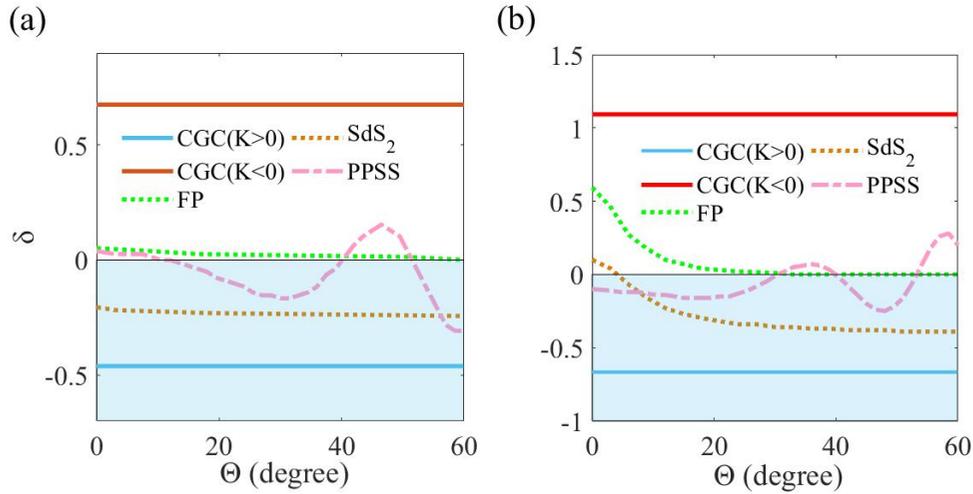

**Fig. 6.** The dependence of the relative quantity $\delta$ on the propagation direction at different propagation distances (a) $D$=400 mm and (b) $D$=500 mm. Other parameters are the same as in Figs. 4 and 5.

## 4. Conclusion

In summary, we have investigated the effect on a single-slit diffraction or double-silt interference of light of curved space with constant and variable Gaussian curvature. The result demonstrates that the width of the main (central) fringe (indicating the optical imaging resolution) in diffraction patterns becomes narrower when Gaussian curvature is positive, and wider when Gaussian curvature is negative, compared with that in flat space, which gives a new possibility to improve the imaging resolution using curved surface with positive curvature.

Using the transformation from curved to flat plane[35,36], one can use 2D plane waveguides with equivalent refractive index profile to improve image resolution. Moreover, the diffraction limit may vary when the spatial curvature of the SORs is no longer a constant, and it may oscillate when the curvature alternates between positive and negative values, like on the SORs of PPSS. The effect increases when the magnitude of Gaussian curvature is larger and the propagation distance is longer. We also show that the resolution keeps constant for arbitrary propagation direction of light when the spatial curvature is constant, and it varies as the local spatial curvature changes. This tells us that the uniformity of space may be reflected by the change of the diffraction limit. Our results will help in the control of optical resolution on curved surfaces and may also probe the non-uniform spacetime in universe space, or in a more realistic perspective, one can expect a method to realize the optical super-resolution in planar integrated waveguide structure in the principle of transformation optics. Moreover, our results may also provide inspiration for electron optics[45,46] extending into synthetic curved spaces[47] or non-Euclidean surfaces[48].

## Acknowledgements

National Natural Science Foundation of China (NSFC) (grants No.11974309); the Israel Science Foundation (No. 1871/15, 2074/15 and 2630/20), the United States-Israel Binational Science Foundation NSF/BSF (No. 2015694).

## Conflict of Interest

The authors declare no conflicts of interest.

## Data availability

The data that support the findings of this study are available from the corresponding author upon reasonable request.